\documentclass[12pt]{article}
\usepackage[dvips]{graphicx}
\usepackage{authblk}

\begin{document}

\title{\hbox{\normalsize UDC 523.92+523.942-337}Parity fluctuations in stellar dynamos}

\author[1]{D.~L.~Moss}
\author[2,3,4]{D.~D.~Sokoloff}
\affil[1]{School of Mathematics, University of Manchester}
\affil[2]{IZMIRAN, Troitsk, Moscow, Russia}
\affil[3]{Moscow State University, Moscow, Russia}
\affil[4]{ICMM, Perm, Russia}

\maketitle

\begin{abstract}

Observations of the solar butterfly diagram from sunspot records
suggest persistent fluctuation in parity, away from the overall, approximately dipolar structure. We use a simple mean-field dynamo model with a solar-like rotation law, and perturb the $\alpha$-effect. We find that the parity of the magnetic
field with respect to the rotational equator can demonstrate  what we describe as
resonant behaviour, while the magnetic energy behaves in a  more or less
expected way. We discuss possible applications of the phenomena in the context of
various deviations of the solar magnetic field from dipolar symmetry,
as reported from analysis of archival sunspot data.
We deduce that our model produces fluctuations in field parity, and hence in the butterfly diagram, that are consistent with observed fluctaions in solar behaviour.

\end{abstract}

\section{Introduction}

The contemporary symmetry of the solar magnetic field is close to dipolar, i.e. the large-scale magnetic 
field is antisymmetric with respect to the solar equator.  Historical telescopic observations 
indicate however that sometimes the solar magnetic field can deviate substantially from this symmetry. This
happened at the end of the Maunder minimum (Ribes \& Nesme-Ribes 1993),
and less clear evidence is available for the 
epoch just before the Maunder minimum (Nesme-Ribes et al. 1994). In each case an activity wave propagated in just 
one (Southern) solar hemisphere,
in what can be described as simultaneous excitation of dipolar and quadrupolar 
magnetic configurations with comparable field strengths (Sokoloff \& Nesme-Ribes 1994). Indeed, spherical dynamo 
models are known to be able to excite magnetic configurations of either symmetry
(e.g. Brandenburg et al. 1989). Recent investigations 
of sunspot drawings from the XVIIIth century (Arlt at al. 2013) give a hint that even an almost pure quadrupolar magnetic 
configuration is sometimes excited (Sokoloff et al, 2010). Some quantities based on sunspot data indicate 
substantial deviations from dipolar symmetry even in the XXth century (Lisseu et al. 2016). 

On one hand, spherical dynamo models allow excitation of dipolar and quadrupolar configurations in addition to what is  known as mixed 
parity configurations (e.g. Moss et al. 2008a). Models which include stochastic fluctuations of dynamo drivers, the $\alpha$-effect primarily,
can mimic episodes with deviations from dipolar symmetry such as are known from 
observations in form of fluctuations of magnetic field parity (e.g. Moss et al. 2008b; Usoskin et al. 2009). 

On the other hand however these parity fluctuations appear as rather rare cases in the parametric space of 
spherical dynamo models, which mainly result in the excitation of magnetic configurations of a particular polarity.       
A feeling that dynamo driver fluctuations associated with a coincidence of some characteristics of dynamo 
excited modes underly the fluctuations in parity follows from the investigations cited above. 
This expectation 
appears quite specific to the dynamo problem.  However it perhaps has a similarity with certain resonant phenomena, and 
for the sake of definiteness we refer to it here as stochastic resonance.

Several attempts have been undertaken to isolate resonant effects in spherical
shell dynamos (Gilman \& Dikpati 2011; Moss 
\& Sokoloff 2013), and also in disc dynamos (Kuzanyan \& Sokoloff 1993;  Moss 1996),
that are relevant for the generation and maintenance of galactic magnetic fields.
These papers however concentrated attention on enhancement of magnetic energy due to the resonant effects, and in general their results are interesting but
are limited in applicability.

Here we readdress the problem of phenomena with substantial fluctuations of parity, that occur in a 
rather small region of parameter space, and are associated with stochastic variations of the dynamo drivers 
(although for comparison we do also briefly consider periodic variations). 
The important point is that we 
focus on
stochastic rather than periodic variations of the dynamo drivers.

As we expect that resonant phenomena are rather general, we consider 
parity fluctuations in the framework of
a simple mean-field dynamo model, with the aim of clarifying 
the conditions associated with parity fluctuations in our models. 
This is the aim of this paper.

A dynamo operating in a spherical shell, driven by mirror asymmetric convection and differential rotation, is believed to be the 
physical process that underlies the magnetic solar activity cycle (e.g. Stix 2004).
Spherical shell dynamos seem to be responsible for  the
activity cycles observed on various late-type stars (Baliunas et al. 1996).
As resonance seems to be a general physical 
phenomenon relevant for various periodic processes it seems natural
to search for resonant behaviour in spherical shell
dynamos, and possibly to associate various properties of solar and 
stellar cycles with resonant effects (e.g. Gilman \& Dikpati 2011).

\begin{figure}[t]
\includegraphics[width=0.45\textwidth]{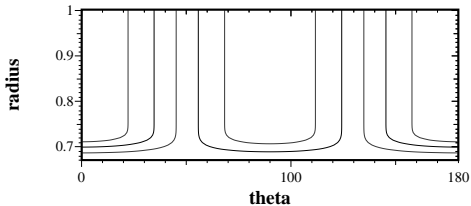}
\caption{The Jouve et al. (2008) rotation law: equally spaced isorotation contours.}
\label{rotn}
\end{figure}

\begin{figure}[t]
\includegraphics[width=0.45\textwidth]{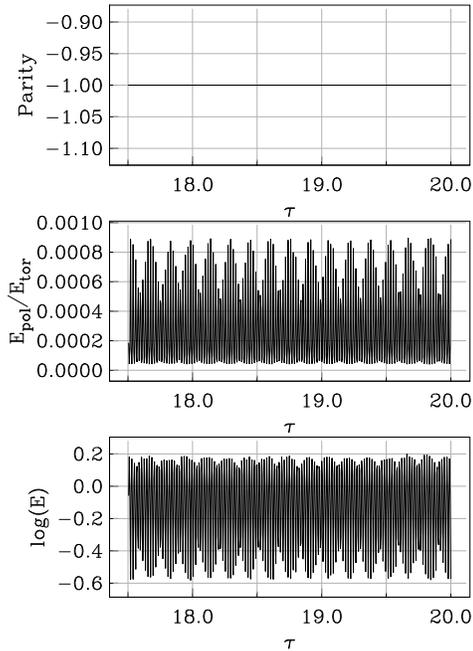}
\caption{Extract from a mature solution showing variations in parity (top), ratio of poloidal to toroidal energies
(middle) and total energy for the model with equatorially symmetric alpha perturbations (case b)).}
\label{ihem=1}
\end{figure}

\section{The dynamo model}
\label{model}

\begin{figure*}
a) \includegraphics[width=0.45\textwidth]{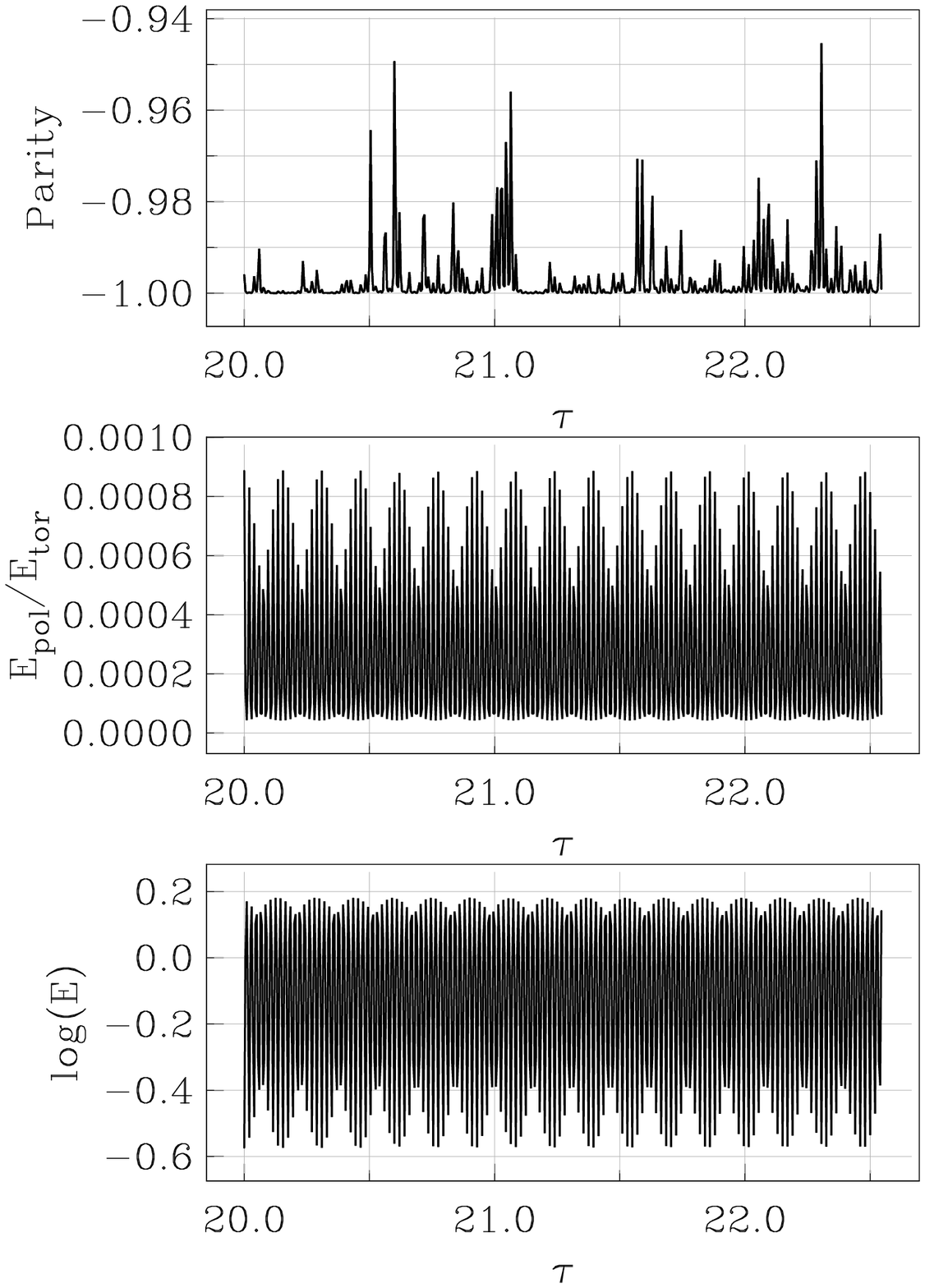}
b) \includegraphics[width=0.45\textwidth]{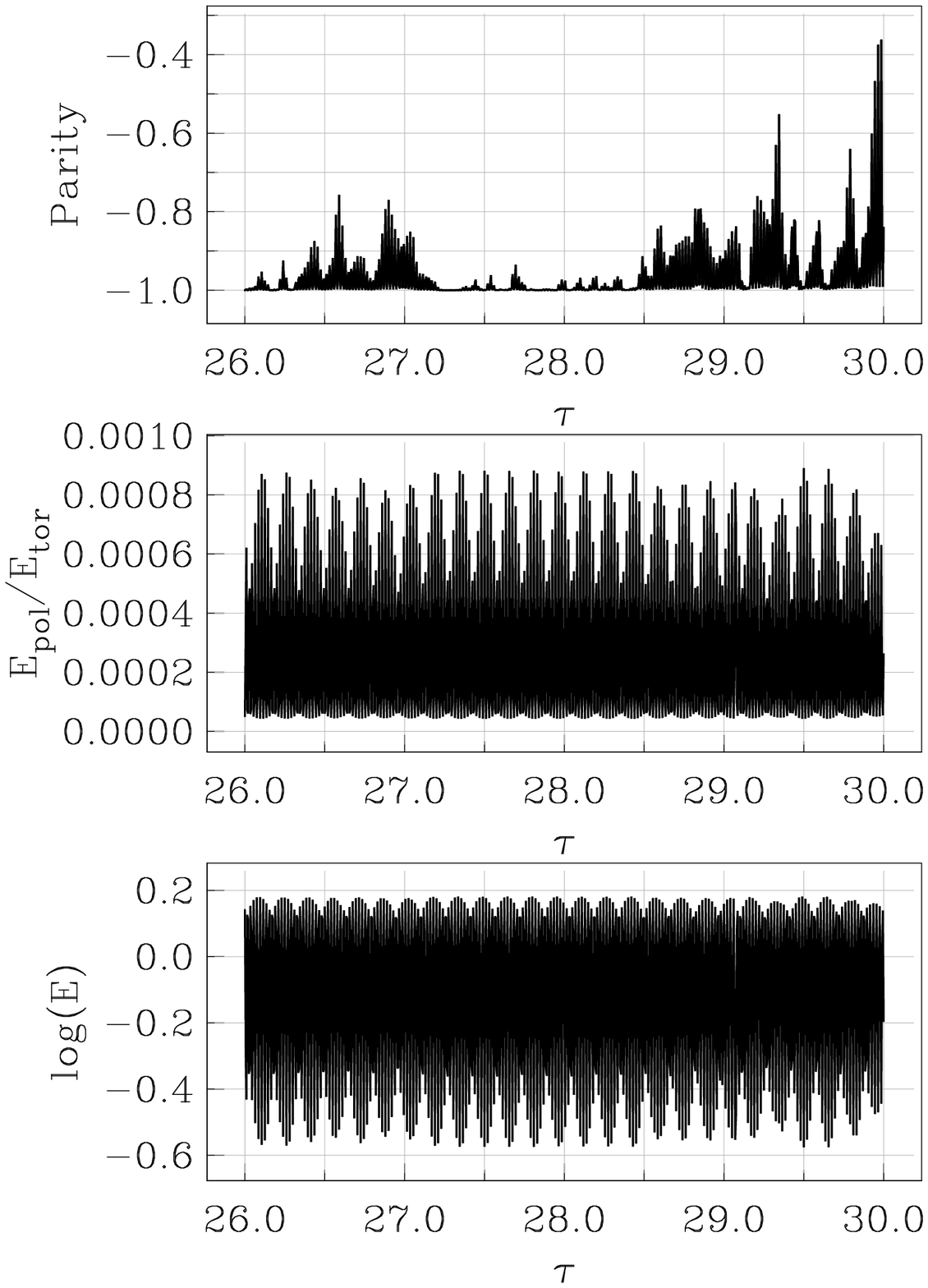}\\
c) \includegraphics[width=0.45\textwidth]{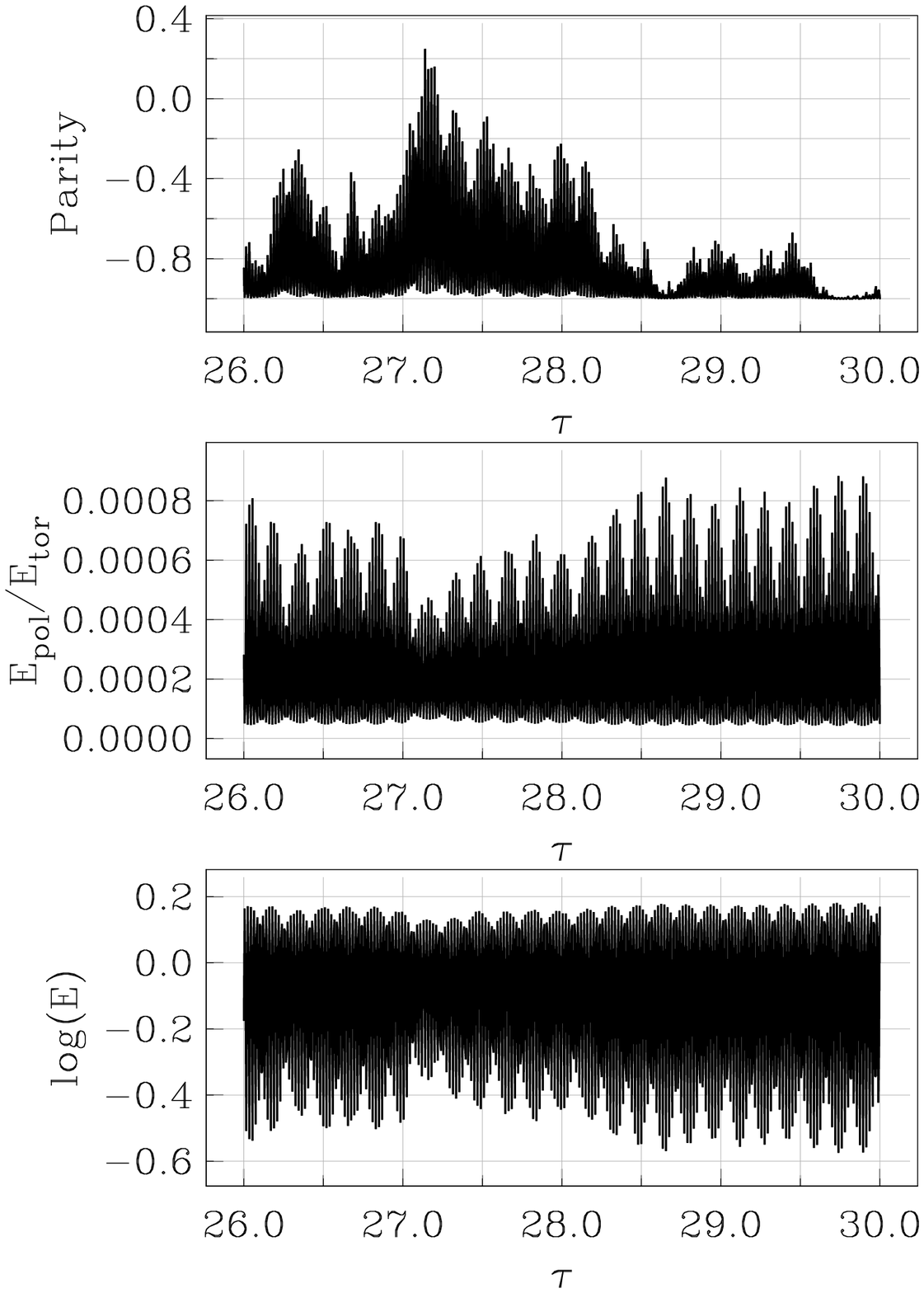}
d) \includegraphics[width=0.45\textwidth]{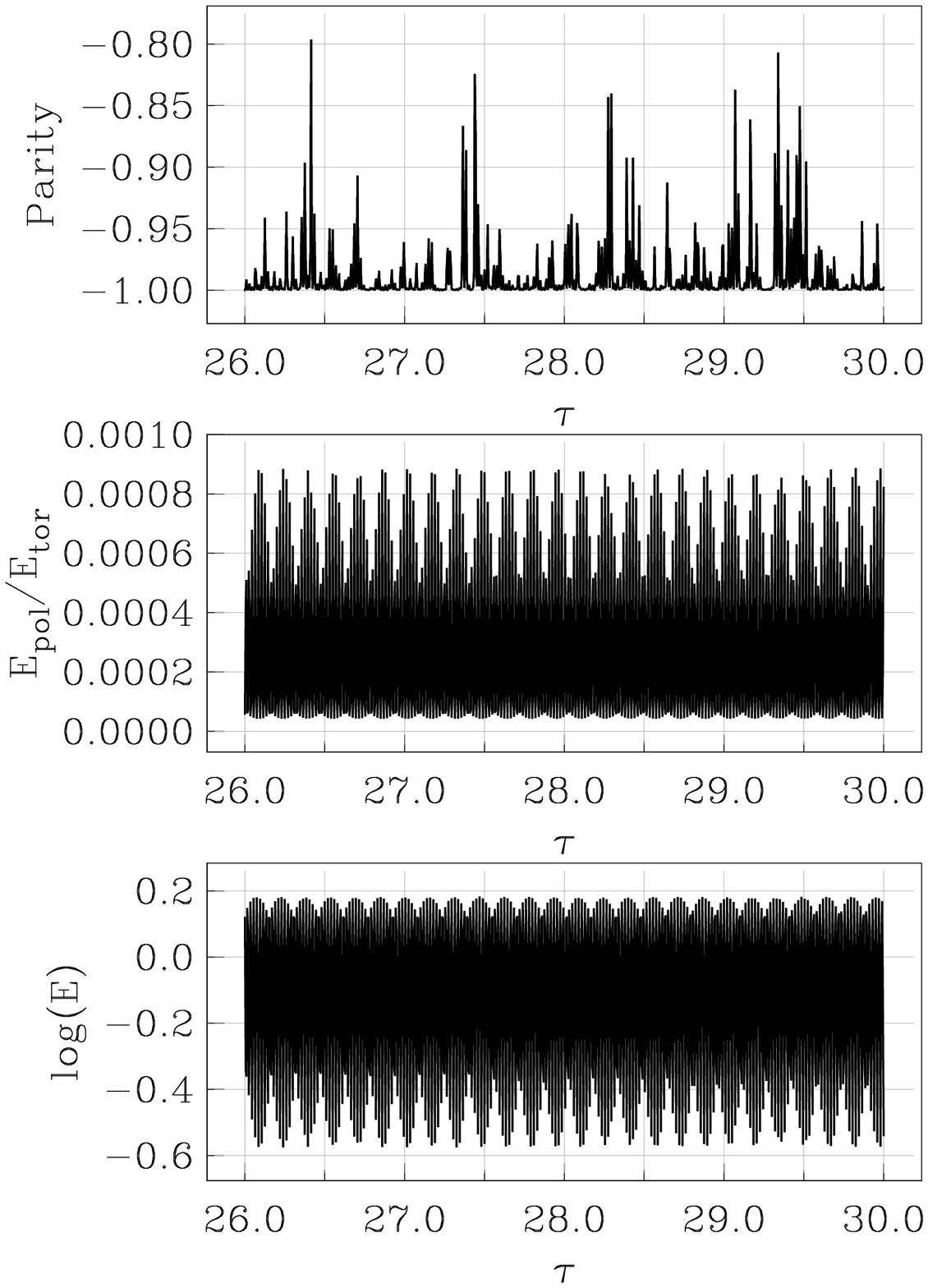}

\caption{Extracts from time series for models with $\alpha$-fluctuations  (s=0.2), all case a). a:  $t_c = 0.019$, $f_r=0.075$, 
b:  $t_c = 0.05$, $f_r=0.075$, c: $t_c=0.005$, $f_r=0.005$, d: $t_c=0.019$,
$f_r=0.150$. The rerandomization time interval $t_c$ is constant.
In each subfigure
the upper panel shows the parity, the middle panel the ratio of global poloidal 
to toroidal energies. and the lower panel gives the  global magnetic energy.}
\label{flucts}
\end{figure*}

We consider a standard mean-field dynamo 
based on the joint action of differential rotation and the $\alpha$-effect,
operating in a spherical shell.
For convenience, we use a synthetic rotation curve, as presented in Jouve et al.
(2008). The isorotation contours are illustrated in Fig.~\ref{rotn}.
Our model extends through a tachocline to solid body rotation at fractional
radius r=0.67.
The diffusivity is taken to be uniform. The boundary condition at the surface 
is that the interior field fits smoothly onto a vacuum exterior field, 
and at the lower boundary we use perfect conductor conditions.
A simple algebraic alpha-quenching is implemented, with further details
such as the unquenched form of alpha given by case A of Jouve et al. (2008).
Salient results are given (NDYND code) in Jouve et al. (2008).
We take a standard dynamo number for our investigations here that is about
twice the supercritical value. The excitation conditions for dipolar and 
quadrupolar modes are well separated, with the dipolar solution strongly preferred.
(To verify this we made several experiments starting with initial fields
of very close to purely even parity: in each case the system evolved rapidly to the same pure odd parity state.)

We then add fluctuations of the dynamo driver $\alpha$ as follows. The unperturbed $\alpha$ is multiplied 
by 
a factor $1+f_r r_i$ where $r_i$ is a sequence of Gaussian random quantities 
(with zero mean and r.m.s. value $s$) that are independent at 
time intervals of fixed duration  $t_c$; $f_r$ is an adjustable scaling factor. 
In one set of models $r_i$ is regenerated periodically at fixed 
intervals $t_c$. 
Then our reference case has standard deviation $s=0.2$, and rerandomizations occur at fixed intervals 
$t_c=0.0075$. 
In another set of computations the interval $t_c$ is replaced by $t_{\rm c0}(1+ s)$ where $s$
is also taken from a Gaussian distribution.
For comparison we consider also periodic $\alpha$-variations with amplitude $r$ and period $t_\alpha$.

\begin{figure}[t]
a) \includegraphics[width=0.45\textwidth]{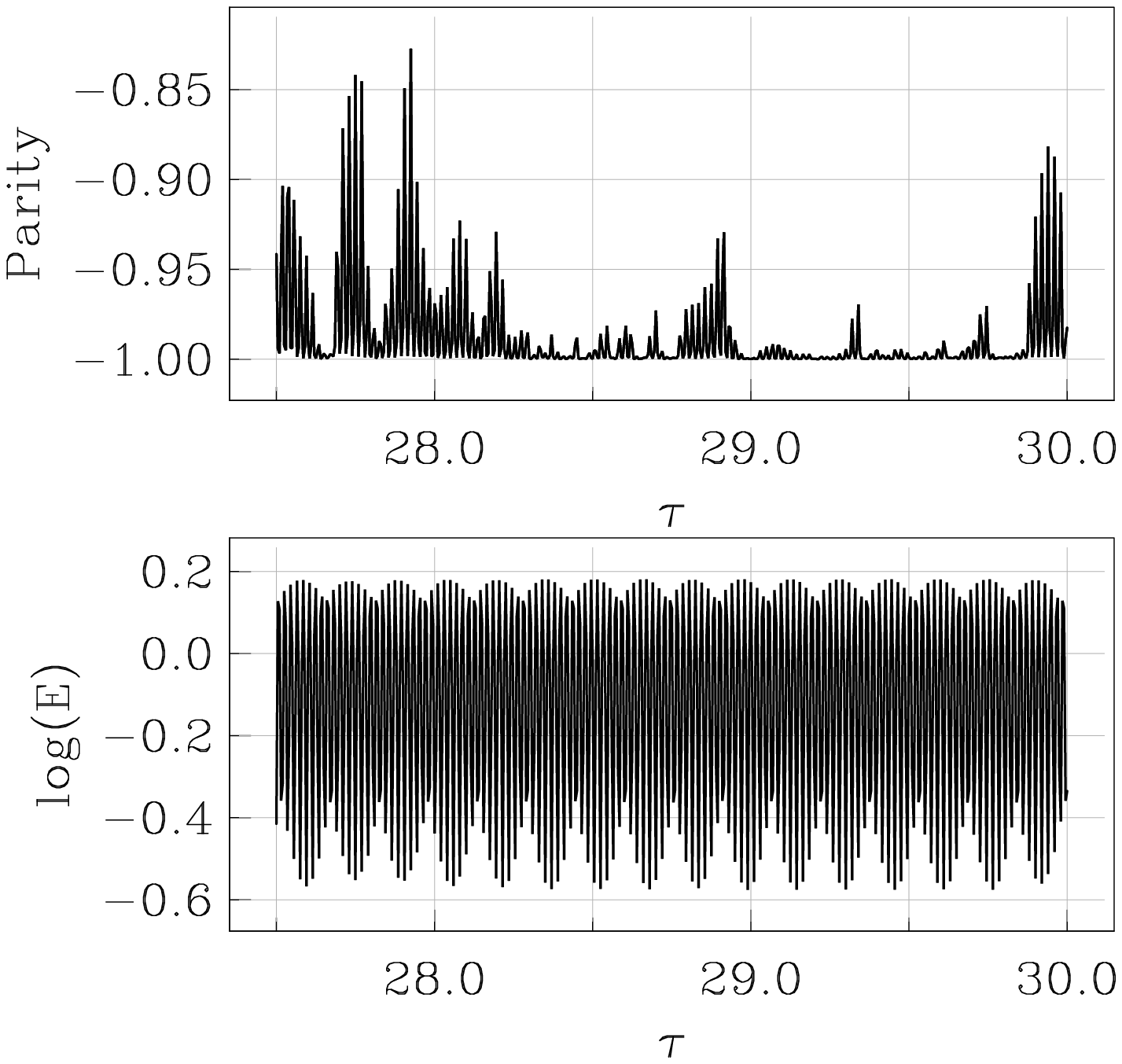}\\
b) \includegraphics[width=0.45\textwidth]{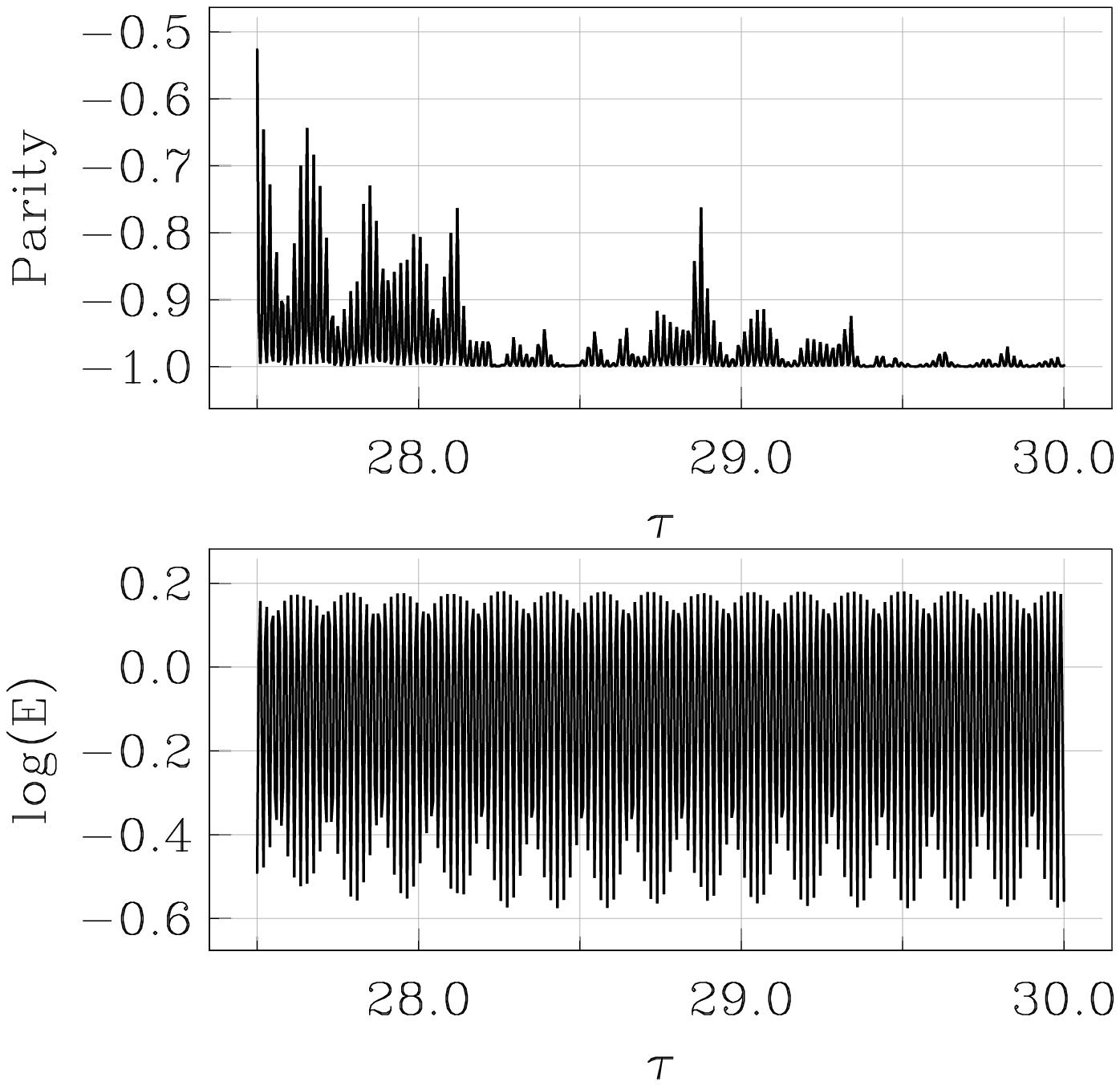}
\caption{Models with fluctuations in $t_c=t_{c0}(1+s)$, where
$t_{c0}=0.0075$ and $s$ has 
mean 0.0.
In panel a), the standard deviation of the interval fluctuations,
determined by $s$, is 0.2, and the other parameters are as in panel a) of Fig.~\ref{flucts}.
In panel b) the standard deviation $s$ of the random interval
is reduced from 0.2 to 0.1.}
\label{randtc}
\end{figure}

\section{Resonant effects}
\label{res}

\subsection{Stochastic $\alpha$-fluctuations}
\label{stoch}

For simplicity and clarity we consider a model in which fluctuations in the alpha
coefficient are applied to each hemisphere as a whole. We consider two possibilities:
a) $r_i(t,N)=-r_i(t,S)$ and b) $r_i(t,N)=r_i(t,S)$,
where  N and S refer to the corresponding hemispheres.
(More general perturbations could, of course, 
be considered, cf Moss et al. 1992.) Our choice is in the spirit of Moss et al. (2007), and in some ways 
can be considered as an extension of that  1D model to  two dimensions.

We find that  that with the antisymmetric perturbations (case a)) the destabilization
of the initial odd parity state is almost immediate, and that the solution
subsequently shows deviations from even parity. 

Case b) with equatorially symmetric perturbations maintains pure parity
with fluctuations in energy  similar to those shown in Fig.~\ref{flucts}a.
In this case, the pure parity state is preserved, and
the magnetic energy shows irregular variations. 
Thus we consider below case a) only.

Fluctuations in parity and energy 
for a selection of mature solutions are shown 
in Fig.~\ref{flucts}.
The form of the parity fluctuations  as well as their amplitude
depend significantly on the fluctuation 
parameters. In particular, the parity fluctuations shown in Fig.~\ref{flucts}c are noticeably stronger 
and longer than in the other cases. 
The magnetic energy and the ratio between 
the energy in toroidal and poloidal magnetic fields do not 
demonstrate any unusual behaviour that could be attributed to resonant effects.
Here the results 
are similar to those obtained previously by Moss \& Sokoloff (2013) 
earlier. The novelty is
in the unusual behaviour of the parity, which
appears chaotic if not quasiperiodic.

We further investigated models in which the injection rerandomization interval 
$t_c$ also varies randomly. In Fig.~\ref{randtc}a we show an extract from 
the mature solution obtained with the same parameters as Fig.~\ref{flucts}a but with $t_c$ replaced by 
$t_{\rm c0}(1+s)$ where s is taken from a Gaussian distribution
with mean 0.0 and standard deviation 0.2, and with $t_{\rm c0}$ the reference time interval (Fig.~\ref{randtc}b).
Reducing the standard deviation of the variable $s$ from 0.2 to 0.1 we do not see substantial 
reorganization of fluctuations observed.

\subsection{Periodic $\alpha$-perturbations}
\label{perpert}

Just for orientation, we also
performed similar experiments with models in which the variations in $\alpha$-fluctuations were periodic. We agree immediately that there is no obvious physical application,
but felt that such experiments might provide some general insight.
Now there is no random input into the models, but $\alpha$ varies with 
period $t_\alpha$. With the same parameters otherwise as in the 
models of Sect.~\ref{stoch} we increased $t_\alpha$ from  about
half of the unperturbed period of the model to about 4 times this value,
again
starting from an odd parity seed field.

Solutions maintain pure parity $P=-1$, but expectedly show fluctuations in total energy.  We found no evidence for resonant behaviour in these examples.
(It is possible that in a model in which excitation conditions for dipolar
and quadrupolar fields are close, the behaviour in this case could be more complex.)

\section{Discussion and conclusions}
\label{disc}

Summarizing, we found
the cases studied show strong fluctuations of parity, which 
appear as chaotic
or even quasi-periodic, in contrast to the fluctuations in energy which
do not display any such remarkable characteristics. Fluctuations of parity becomes especially pronounced, i.e. large amplitude and time scale, 
for particular tunings of the dynamo governing parameters (Fig.~3c) and in this sense we are speaking about a resonant effect. The parity fluctuations found 
are a consequence of the  stochastic fluctuations of the dynamo drivers. In this sense we refer the case as stochastic resonance.   

As emphasized, we note that substantial deviations from dipolar symmetry have been reported 
for the solar magnetic field at the end of 
the Maunder minimum (Ribes \& Nesme-Ribes 1993; Sokoloff \& Nesme-Ribes 1994), 
at the times just before the Maunder 
minimum (Nesme-Ribes et al. 1994), and in 
the epoch following the Maunder minimum (Sokoloff et al. 2010). 
In these cases, a tracer of parity 
demonstrates a significant deviation from dipolar parity. 
Substantial deviations from dipolar parity are claimed
even for  a cycle 
at the end of the XXth century 
(Leussu et al. 2016).  Of course, the reliability of archival observations is 
necessarily lower than for 
contemporary data (Zolotova \& Ponyavin 2015), and even the detailed 
reliability of sunspot data in the XXth century is currently a 
matter for discussion (e.g. Clette et al. 2014). However unusual behaviour of solar activity 
as recorded in parity data is accepted for  the time being at least as
lying within  the mainstream of
solar physics (Usoskin et al. 2015). Stochastic resonance similar to that investigated here should 
be considered as a possible physical effect underlying the above phenomenology.

Further, fluctuations of dynamo drivers are considered as the most probable factor 
that underlies Grand minima and other related phenomena in solar activity 
(e.g. Moss et al. 2008;  Olemskoy et al. 2013).
It seems  plausible to suggest that resonant excitation of parity variations by random fluctuations of 
dynamo drivers could be a viable addition to the understanding of the 
Maunder minimum. We note here that the resonant nature 
of such phenomena was recently discussed by Stefani et al.(2016) 

DS acknowledges financial support from the RSF under grant 16-41-02012.

\end{document}